\documentclass{elsart}

\usepackage{graphicx}
\usepackage{bm}

\newcommand{\gdhi}{\ooalign{\hfil/\hfil\crcr$\partial$}}

\begin{document}
\begin{frontmatter}


\title{Degeneracy of the quarks, shell structure in the chiral soliton}

\author{Nobuyuki Sawado},
\ead{sawado@ph.noda.tus.ac.jp}
\author{Noriko Shiiki}
\ead{norikoshiiki@mail.goo.ne.jp}
\address{Department of Physics, Tokyo University of Science, Noda, Chiba 278-8510, Japan}

\date{\today}

\begin{abstract}
We obtain multi-soliton solutions with discrete symmetries in the chiral quark 
soliton model using the rational map ansatz. The solutions exhibit degenerate 
bound spectra of the quark orbits depending on the background of pion field configurations. 
It is shown that resultant baryon densities inherit the same discrete symmetries as 
the chiral fields. Evaluating the radial component of the baryon density, 
shell-like structure of the valence quark spectra is also observed. 
\end{abstract}
\end{frontmatter}

\section{\label{sec:level1}Introduction}
The chiral quark soliton model (CQSM) was developed in 
1980's as a low-energy effective theory of QCD. 
Since it includes the Dirac sea quark 
contribution and explicit valence quark degrees of 
freedom, the model interpolates between the 
constituent quark model and the Skyrme 
model \cite{diakonov88,reinhardt88,meissner89,report,wakamatsu91}. 
The CQSM incorporates the non-perturbative
feature of the low-energy QCD, spontaneous chiral symmetry breaking. 
It has been shown that the $B=1$ solution provides correct observables 
such as a nucleon including mass, electromagnetic value, spin carried
by quarks, parton distributions and octet
SU(3) baryon spectra. 
For $B=2$, the stable axially symmetric
soliton solution was found in Ref.~\cite{sawado98}. 
The solution exhibits doubly degenerate bound spectrum 
of the quark orbits in the background of axially symmetric  
chiral fields with winding number two. Upon quantization, 
various dibaryon spectra were obtained, showing that the quantum 
numbers of the ground state coincide with those of physical 
deuteron~\cite{sawado00,sawado02}. For $B>2$, the Skyrme model 
predicts that minimum energy solutions have discrete, 
crystal-like symmetries~\cite{braaten90,sutcliffe97,manton98}. 
According to this prediction, we studied the CQSM with $B=3$ 
tetrahedrally symmetric chiral fields and obtained triply 
degenerate spectrum of the quark orbits~\cite{sawado02t}. 
Its large degeneracy indicates that the tetrahedrally symmetric 
solution may be the lowest-lying configuration. 
Thus, for $B>3$, one can also expect that the lowest-lying solutions 
in the CQSM inherits the discrete symmetries predicted in the Skyrme model
and have large degeneracy. 

In this paper we shall study soliton solutions with $B\ge 3$ 
in the CQSM using the rational map ansatz for the background of pion fields. 
The procedure of the numerical computation performed here 
is shown in Sec.\ref{sec:level3}. 
In Sec.\ref{sec:level4}, we show obtained classical 
self-consistent soliton solutions with $B=3-9$. 
These solutions exhibit various degenerate spectra of the 
quark orbits depending on the symmetry of the background 
chiral fields. Such degeneracy generates large shell gaps, which suggests 
that the solutions are stable local minima.  
Evaluating the radial component of the baryon density, shell-like structure 
of the valence quarks can be observed.  
The results and discussion are summarized in Sec.\ref{sec:level5}.  
 
\section{\label{sec:level2}The model}
The CQSM incorporates the nonperturbative feature of the low-energy QCD, 
spontaneous chiral symmetry breaking with the vacuum functional defined by \cite{diakonov88}
\begin{eqnarray}
	{\mathcal Z} = \int {\mathcal D}\pi{\mathcal D}\psi{\mathcal D}
	\psi^{\dagger}\exp \left[ 
	i \int d^{4}x \, \bar{\psi} \left(i\!\!\not\!\partial
	- MU^{\gamma_{5}}\right) \psi \right],	 \label{vacuum_functional}
\end{eqnarray} 
where the SU(2) matrix 
\begin{eqnarray}
	U^{\gamma_{5}}= \frac{1+\gamma_{5}}{2} U + \frac{1-\gamma_{5}}{2} U^{\dagger} 
	\,\,\,{\rm with} \,\,\,\,
	U=\exp \left( i \bm{\tau} \!\cdot\! \bm{\pi}/f_{\pi} \right) \nonumber
\end{eqnarray}
describes chiral fields, $\psi$ is quark fields and $M$ is the constituent 
quark mass. $f_{\pi} $ is the pion decay constant and experimentally 
$f_{\pi} \sim 93 {\rm MeV}$. 

The $B=1$ soliton solution has been studied in detail at classical and 
quantum level in Refs.~\cite{diakonov88,reinhardt88,meissner89,report,wakamatsu91}.  
To obtain solutions with $B>1$, we shall employ the chiral field configuration with 
winding number $B$ obtained in the Skyrme model as the background of quarks, 
which can be justified as follows. 
In Eq.~(\ref{vacuum_functional}), performing the functional integral over 
$\psi$ and $\psi^\dagger$ fields, one obtains the effective action
\begin{equation}
	S_{\rm eff}[U]=-iN_{c} {\rm Sp} \ln iD = -iN_c \log \det iD, 
	\label{effective_action2}
\end{equation}
where $iD=i \gdhi - M U^{\gamma_5}$ is the Dirac operator.
The classical solutions can be derived by imposing an extremum condition 
on the effective action with respect to $U$. 
For this purpose, let us consider the derivative expansion of 
the action \cite{wakamatsu91,dhar85,ebert86}. 
Up to quartic terms, we have  
\begin{eqnarray}
	S_{\rm eff}&=&\int d^4x \biggl[-C {\rm tr}(L_\mu L^\mu)
	\nonumber \\
	&+&\frac{N_c}{32\pi^2}
	{\rm tr}\Bigl \{\frac{1}{12}[L_\mu,L_\nu]^2-\frac{1}{3}(\partial_\mu L^\mu)^2
	+\frac{1}{6}(L_\mu L^\mu)^2 \Bigr \} \biggr], 
      \label{expand_action}
\end{eqnarray}
where $L_\mu=U^\dagger \partial_\mu U$. Suitably adjusting the coefficients, 
one can identify the first two terms of Eq.~(\ref{expand_action}) with the Skyrme 
model action. However, the 4th order terms tend to distabilize solutions and  
no stable classical solution can be obtained from the above action~\cite{dhar85,aitchison85}. 
Nevertheless, because of their similarity, it will be justified to adopt the 
configurations of the solutions in the Skyrme model to chiral fields in the CQSM. 

In the CQSM, the number of valence quark is associated with 
the baryon number such that a soliton with baryon number $B$ 
consists of $N_c\times B$ valence quarks. 
If the quarks are strongly bound inside the soliton, 
their binding energy become large and the valence quarks 
can not be observed as positive energy particles~\cite{kahana84,balachandran98}.
Thus, one gets the picture of the topological soliton model 
in the sense that the baryon number coincides with the winding 
number of the background chiral field when the valence quarks 
occupy all the levels diving into negative energy region. 

Let us rewrite the effective action in Eq. (\ref{effective_action2}) as 
\begin{eqnarray}
	S_{\rm eff}\to	
	-iN_c\log \det\bigl(i\partial_t-H(U^{\gamma_5})\bigr) \label{effective_det},
\end{eqnarray}
and introduce the eigenequation 
\begin{eqnarray}
	&& H(U^{\gamma_5})\phi_\mu(\bm{x})=E_\mu\phi_\mu(\bm{x})\,, \\
	\label{eigeneq} 
	&& H(U^{\gamma_5})=-i\alpha\cdot\nabla + \beta MU^{\gamma_{5}}\,.
	\label{hamiltonian}
\end{eqnarray}
The effective action $S_{\rm eff}(U)$ is ultraviolet divergent 
and hence must be regularized. Using the proper-time regularization scheme~\cite{schwinger51}, 
one can write 
\begin{eqnarray}
	&&S^{{\rm reg}}_{{\rm eff}}[U]
	=\frac{i}{2}N_{c}
	\int^{\infty}_{1/\Lambda^2}\frac{d\tau}{\tau}{\rm Sp}\left(
	{\rm e}^{-D^{\dagger}D\tau}-{\rm e}^{-D_{0}^{\dagger}D_{0}\tau}\right) 
	\nonumber \\
	&&\hspace{12mm}=\frac{i}{2}N_{c}T\int^{\infty}_{-\infty}\frac{d\omega}{2\pi}
	\int^{\infty}_{1/\Lambda^2}\frac{d\tau}{\tau}
	{\rm Sp}\Bigl[{\rm e}^{-\tau (H^2+\omega^2)} 
	-{\rm e}^{-\tau (H_{0}^2+\omega^2)}\Bigr],
	\label{regularized_action}
\end{eqnarray}
where $T$ is the Euclidean time separation, $D_{0}$ and $H_{0}$ are operators with $U=1$.

At $T \rightarrow \infty$, we have ${\rm e}^{iS_{{\rm eff}}}
\sim  {\rm e}^{-iE_{\rm field}T}$. 
The total energy is then given by
\begin{equation}
	E_{\rm static}[U]=E_{\rm val}[U]+E_{\rm field}[U]-E_{\rm field}[U=1],
	\label{total_energy}
\end{equation}
where 
\begin{eqnarray}
	E_{\rm val}=N_c\sum_{i}E^{(i)}_{\rm val} 
\end{eqnarray}
is the valence quark contribution with the valence energy 
$E^{(i)}_{\rm val}$ of the $i$ th valence quark, and the vacuum part is
\begin{eqnarray}
	E_{\rm field}=N_c\sum_{\mu}\left\{ {\mathcal N}(E_\mu)|E_{\mu}|+\frac{\Lambda}
	{\sqrt{4\pi}}\exp \left[ - \left( \frac{E_{\mu}}{\Lambda} \right)^2 
	\right] \right\},
\end{eqnarray}
with
\begin{eqnarray}
	{\mathcal N}(E_{\mu})= -\frac{1}{\sqrt{4\pi}}\Gamma \left(\frac{1}{2}, 
	\left(\frac{E_{\mu}}{\Lambda} \right)^2 \right)\,. \nonumber 
\end{eqnarray}
$\Lambda$ is a cutoff parameter evaluated by the condition that the 
derivative expansion of Eq.~(\ref{regularized_action}) reproduces 
the pion kinetic term with the correct coefficient,

\begin{eqnarray}
	f_{\pi}^2=\frac{N_{c}M^2}{4\pi^2}\int^{\infty}_{1/\Lambda^2} 
	\frac{d\tau}{\tau}{\rm e}^{-\tau M^2}
	\, . \label{cutoff_parameter}
\end{eqnarray}
The contribution to the total energy in the absence of the chiral fields
($U=1$) can be estimated using the eigenstates of the eigenequation, 
\begin{eqnarray}
&&H_0\phi_\mu^{(0)}(\bm{x})=E_\mu^{(0)}\phi_\mu^{(0)}(\bm{x})\,,\\
&&H_0=-i\alpha\cdot\nabla + \beta M\,.
\label{eigen0_h}
\end{eqnarray}

In the Skyrme model it is known that solitons with $B\ge 3$ have 
discrete crystal-like symmetries~\cite{braaten90}. Therefore, we expect that  
soliton solutions of the CQSM inherits the same discrete symmetry as skyrmions. 
However, it is too complicated to perform a numerical computation if one 
imposes such discrete symmetries directly on the chiral fields. Thus  
Houghton, Manton and Sutcliffe proposed remarkable ansatz for the chiral fields, 
rational map ansatz~\cite{manton98}. 
According to this ansatz, the chiral fields are expressed in a rational map as 
\begin{eqnarray}
U(r,z)=\exp(i F(r) \hat{\bm{n}}_R\cdot \bm{\tau})\,,
\label{chiral_fields}
\end{eqnarray}
where
\begin{eqnarray}
\hat{\bm{n}}_R=\frac{1}{1+|R(z)|^2}(2{\rm Re}[R(z)],2{\rm Im}[R(z)],1-|R(z)|^2)
\nonumber
\end{eqnarray}
and $R(z)$ is the rational map. The complex coordinate $z$ is given by  
$z=\tan (\theta/2)e^{i\varphi}$ via stereographic projection. 

Rational maps are maps from $CP(1)$ to $CP(1)$ (equivalently, from $S^2$ 
to $S^2$) classified by winding number. In Ref.~\cite{manton98} Manton {\it et al.} 
showed that $B=N$ skyrmions can be well-approximated by rational maps with 
winding number $N$. The rational map with winding number $N$ possesses 
$(2N+1)$ complex parameters whose values can be determined by 
imposing the symmetry of the skyrmion. We shall use this ansatz for 
the background chiral fields in the CQSM. 
Their explicit forms are presented in Appendix \ref{sec:levela}.
Since the chiral fields in Eq.~(\ref{chiral_fields}) is parameterized by polar coordinates, 
one can apply the numerical technique developed for $B=1$ to find solutions with higher $B$.
Detailed calculations will be presented in the subsequent section.  

Field equations for the chiral fields can be obtained by demanding 
that the total energy in Eq.~(\ref{total_energy}) be stationary 
with respect to variation of the profile function $F(r)$,
\begin{eqnarray*}
	\frac{\delta}{\delta F(r)}E_{\rm static}=0 \,\, ,
\end{eqnarray*}
which produces  
\begin{eqnarray}
	S(r)\sin F(r)=P(r)\cos F(r),  
	\label{field_equation}
\end{eqnarray}
where 
\begin{eqnarray}
&&S(r)=N_{c}\sum_\mu\bigl(n_\mu\theta(E_\mu)+{\rm sign}(E_\mu)
{\mathcal N}(E_\mu)\bigr)
\langle \mu |\gamma^{0}\delta(|x|-r)|\mu\rangle\,, 
\\	
&&P(r)=N_{c}\sum_\mu\bigl(n_\mu\theta(E_\mu)+{\rm sign}(E_\mu)
{\mathcal N}(E_\mu)\bigr)
\langle \mu |i \gamma^{0}\gamma^{5}\hat{\bm{n}}_{R}
\cdot\bm{\tau}\delta(|x|-r)|\mu\rangle \, .\nonumber \\
\end{eqnarray}
The procedure to obtain self-consistent solutions of Eq.~(\ref{field_equation}) 
is that $1)$ solve the eigenequation in Eq. (\ref{eigeneq}) under an assumed 
initial profile function $F_{0}(r)$, $2)$ use the resultant eigenfunctions and 
eigenvalues to calculate $S(r)$ and $P(r)$, $3)$ solve 
Eq. (\ref{field_equation}) to obtain a new profile function, $4)$ repeat $1)-3)$ 
until the self-consistency is attained.

The baryon density $b(\bm{x})$ is defined by the zeroth component 
of the baryon current~\cite{reinhardt88}; 
\begin{eqnarray}
	&&b(\bm{x}) = \frac{1}{N_{c}}\langle\bar{\psi}\gamma_{0}
	\psi\rangle=b_{\rm val}(\bm{x})+b_{\rm field}(\bm{x}),
\label{baryon_density}
\end{eqnarray}
where
\begin{eqnarray}
	&&b_{\rm val}(\bm{x})= \sum_i b_{\rm val}^{(i)}(\bm{x})
	=\frac{1}{N_{c}}
	\sum_{i}  
	\phi_i(\bm{x})^\dagger \phi_i(\bm{x})\,,
	\nonumber \\
	&&b_{\rm field}(\bm{x})=\frac{1}{N_c}\Bigl
	[\sum_{\mu}{\rm sign}(E_\mu){\mathcal N}(E_\mu)	
	\phi_\mu(\bm{x})^\dagger \phi_\mu(\bm{x})\nonumber \\
	&&\hspace{1cm}-\sum_{\mu}{\rm sign}(E_\mu^{(0)}){\mathcal N}(E_\mu^{(0)})	
	\phi_\mu^{(0)}(\bm{x})^\dagger \phi_\mu^{(0)}(\bm{x})
	\Bigr]\,.
	\label{baryon_density_element}
\end{eqnarray}

To examine the shell structure of the quarks, we evaluate the radial 
density for the $i$th valence quark $\rho^{(i)}(r)$ in which 
the angular degrees of freedom are integrated out, via, 
\begin{eqnarray}
\rho^{(i)}(r)=\int d\varphi\int\sin\theta d\theta~b^{(i)}_{\rm val}
(r,\theta,\varphi)
\end{eqnarray}
with the baryon number 
\begin{eqnarray}
B=\sum_i \int dr r^2\rho^{(i)}(r)\,.
\end{eqnarray}

\section{\label{sec:level3}Numerical technique}
The numerical method that is widely used in this model is 
based on the expansion of the Dirac spinor in the appropriate 
orthogonal basis. However, since the symmetries with $B\ge 3$ 
are discrete and it is hard to find any symmetric operator commuting 
with the hamiltonian, we shall expand the Dirac field in the Kahana-Ripka 
basis as in the case of $B=1$~\cite{kahana84}. 
The Kahana-Ripka basis which was originally constructed for diagonalizing the 
hamiltonian with the chiral fields of $B=1$ hedgehog ansatz is a plane-wave 
finite basis. The basis is discretized by imposing an appropriate boundary 
condition on the radial wave functions at the radius $r_{\rm max}$ chosen 
to be sufficiently larger than the soliton size. 
The basis is then made finite by including only those states 
with the momentum $k$ as $k<k_{\rm max}$. 
The results should be, however, independent on $r_{\rm max}$ and $k_{\rm max}$.

The hamiltonian with hedgehog ansatz commutes with the parity and the grandspin operator 
given by  
\begin{eqnarray*}
	\bm{K}=\bm{j}+\bm{\tau}/2=\bm{l}+\bm{\sigma}/2+\bm{\tau}/2,
\end{eqnarray*}
where $\bm{j},\bm{l}$ are respectively total angular momentum and orbital angular momentum. 
Accordingly, the angular basis can be written as  
\begin{eqnarray}
|(lj)KM\rangle= \sum_{j_3\tau_3}C^{KM}_{jj_3\frac{1}{2}\tau_3}
\Bigl(\sum_{m\sigma_3}C^{jj_3}_{lm\frac{1}{2}\sigma_3}
|lm \rangle |\frac{1}{2}\sigma_3 \rangle \Bigr) |\frac{1}{2} \tau_3 \rangle\,.
\end{eqnarray}
With this angular basis, the normalized eigenstates of the free hamiltonian 
in a spherical box with radius $r_{\rm max}$ can be constructed as follows:
\begin{eqnarray}
&&u^{(1)}_{KM}=
N_k\left( 
\begin{array}{c}
ij_{K}(kr)|(K K+\frac{1}{2})KM\rangle \\
\Delta_k j_{K+1}(kr)|(K+1 K+\frac{1}{2})KM\rangle
\end{array}
\right), \nonumber \\
&&u^{(2)}_{KM}=
N_k\left( 
\begin{array}{c}
ij_{K}(kr)|(K K-\frac{1}{2})KM\rangle \\
-\Delta_k j_{K-1}(kr)|(K-1 K-\frac{1}{2})KM\rangle
\end{array}
\right), \nonumber \\
&&u^{(3)}_{KM}=
N_k\left( 
\begin{array}{c}
i\Delta_k j_{K}(kr)|(K K+\frac{1}{2})KM\rangle \\
-j_{K+1}(kr)|(K+1 K+\frac{1}{2})KM\rangle
\end{array}
\right), \nonumber \\
&&u^{(4)}_{KM}=
N_k\left( 
\begin{array}{c}
i\Delta_kj_{K}(kr)|(K K-\frac{1}{2})KM\rangle \\
j_{K-1}(kr)|(K-1 K-\frac{1}{2})KM\rangle
\end{array}
\right), \nonumber \\
\nonumber \\
&&v^{(1)}_{KM}=
N_k\left( 
\begin{array}{c}
ij_{K+1}(kr)|(K+1 K+\frac{1}{2})KM\rangle \\
-\Delta_k j_{K}(kr)|(K K+\frac{1}{2})KM\rangle
\end{array}
\right), \nonumber \\
&&v^{(2)}_{KM}=
N_k\left( 
\begin{array}{c}
ij_{K-1}(kr)|(K-1 K-\frac{1}{2})KM\rangle \\
\Delta_k j_{K}(kr)|(K K-\frac{1}{2})KM\rangle
\end{array}
\right), \nonumber \\
&&v^{(3)}_{KM}=
N_k\left( 
\begin{array}{c}
i\Delta_k j_{K+1}(kr)|(K+1 K+\frac{1}{2})KM\rangle \\
j_{K}(kr)|(K K+\frac{1}{2})KM\rangle
\end{array}
\right), \nonumber \\
&&v^{(4)}_{KM}=
N_k\left( 
\begin{array}{c}
i\Delta_kj_{K-1}(kr)|(K-1 K-\frac{1}{2})KM\rangle \\
-j_{K}(kr)|(K K-\frac{1}{2})KM\rangle
\end{array}
\right), \label{kahana_ripka}
\end{eqnarray}
with
\begin{eqnarray}
	N_k=\biggl[\frac{1}{2}r_{\rm max}^3
	\Bigl(j_{K+1}(kr_{\rm max})\Bigr)^2\biggr]^{-1/2}
\end{eqnarray}
and $\Delta_k=k/(E_k+M)$. 

The momenta are discretized by the 
boundary condition $j_K(k_i r_{\rm max})=0$.
The  $u,v$ correspond to the {\it ``natural''} 
and {\it ``unnatural''} components of the basis  
which stand for parity $(-1)^{K}$ and $(-1)^{K+1}$ respectively. 

Let us construct the trial function using the Kahana-Ripka basis 
to solve the eigenequations in Eq.~(\ref{eigeneq}), 
\begin{eqnarray}
	&&\phi_\mu(\bm{x})=\lim_{K_{\rm max}\to\infty}\sum^{K_{\rm max}}_{K=0}
      \sum^{K}_{M=-K}\sum_{j=1}^{4}[\alpha_{KM,\mu}^{(j)}u_{KM}^{(j)}(r,\theta,\varphi)\nonumber \\
      &&\hspace{3cm}+\beta_{KM,\mu}^{(j)}v_{KM}^{(j)}(r,\theta,\varphi)]	
	\label{expand_basis}.
\end{eqnarray}

According to the Rayleigh-Ritz variational method \cite{bransden}, the upper 
bound of the spectrum can be obtained from the secular equation
\begin{eqnarray}
\left|\begin{array}{cc}
	{\mathcal A}^{n,n}-E {\mathcal C}^{n,n} 
	& {\mathcal B}^{n,u} \\
	{\mathcal B}^{u,n}
	& {\mathcal A}^{u,u}-E {\mathcal C}^{u,u}
	\end{array}
	\right|=0 \,,
	\label{secular_equation}
\end{eqnarray}
where 
\begin{eqnarray}
	{\mathcal A}^{n,n}_{j,l}(K^{\prime}M^{\prime}, KM)&=&
	\int d^{3}x {u_{K^{\prime}M^{\prime}}^{(j)}}^\dagger H
	u_{KM}^{(l)}\,, \nonumber \\
	{\mathcal B}^{n,u}_{j,l}(K^{\prime}M^{\prime}, KM)&=&
	\int d^{3}x {u_{K^{\prime}M^{\prime}}^{(j)}}^\dagger H
	v_{KM}^{(l)}\,, \nonumber \\ 
	{\mathcal C}^{n,n}_{j,l}(K^{\prime}M^{\prime}, KM)&=&
	\int d^{3}x {u_{K^{\prime}M^{\prime}}^{(j)}}^\dagger
	u_{KM}^{(l)}\,. \label{secular_comp} 
\end{eqnarray}
${\mathcal A}^{u,u},{\mathcal B}^{u,n},{\mathcal C}^{u,u}$ are given by interchanging 
$u$ and $v$ in Eq.~(\ref{secular_comp}).
For $K_{\rm max} \rightarrow \infty$, the spectrum $E$ becomes exact. 
Eq.~(\ref{secular_equation}) can be solved numerically. 

\begin{table}
\begin{center}
\caption{\label{tab:helement3} A schematic picture of the matrix elements  ${\mathcal A}
(K^{\prime}M^{\prime}, KM)$ for the case of $B=3$ (and $B=5$), 
up to $K,K^{\prime}=2$. $S$, $P_1$, $P_0$ and $P_{-1}$ refer to the elements coupled 
with $(K,M)=(0,0)$, $(1,1)$, $(1,0)$ and $(1,-1)$ respectively. Other elements are all 0.}
\begin{tabular}{cccccccccc} \hline
      & (0 0)& (1 1)& (1 0)& (1-1) &(2 2) & (2 1)& (2 0)& (2-1)& (2-2) \\
\hline
(0 0)& $S$ &       &       &        &      &        & $S$  &       &        \\
(1 1)&     & $P_1$ &       &        &      &        &      & $P_1$ &        \\
(1 0)&     &       & $P_0$ &        & $P_0$&        &      &       & $P_0$  \\
(1-1)&     &       &       &$P_{-1}$&      &$P_{-1}$&      &       &        \\
(2 2)&     &       &$P_0$  &        & $P_0$&        &      &       & $P_0$  \\
(2 1)&     &       &       &$P_{-1}$&      &$P_{-1}$&      &       &        \\
(2 0)& $S$ &       &       &        &      &        & $S$  &       &        \\
(2-1)&     &  $P_1$&       &        &      &        &      &  $P_1$&        \\
(2-2)&     &       & $P_0$ &        & $P_0$&        &      &       &$P_0$   \\
\hline
\end{tabular}
\end{center}
\end{table}

\begin{table}
\begin{center}
\caption{\label{tab:helement_nd3} Matrix elements  ${\mathcal B}^{n,u}
(K^{\prime}M^{\prime}, KM)$ for $B=3$ (and $B=5$) 
up to $K,K^{\prime}=2$. $S$, $P_1$, $P_0$ and $P_{-1}$ refer to the elements coupled 
with $(K,M)=(0,0)$, $(1,1)$, $(1,0)$ and $(1,-1)$ respectively. Other elements are all 0.}
\begin{tabular}{cccccccccc}\hline
      & (0 0)& (1 1)& (1 0)& (1-1) &(2 2) & (2 1)& (2 0)& (2-1)& (2-2) \\
\hline
(0 0)&      &        &  $S$ &      &  $S$ &       &      &        & $S$  \\
(1 1)&      &        &      & $P_1$&      & $P_1$ &      &        &      \\
(1 0)& $P_0$&        &      &      &      &       & $P_0$&        &      \\
(1-1)&      &$P_{-1}$&      &      &      &       &      &$P_{-1}$&      \\
(2 2)&$P_0$ &        &      &      &      &       & $P_0$&        &      \\
(2 1)&      &$P_{-1}$&      &      &      &       &      &$P_{-1}$&      \\
(2 0)&      &        & $S$  &      & $S$  &       &      &        & $S$  \\        
(2-1)&      &        &      & $P_1$&      &$P_1$  &      &        &      \\
(2-2)& $P_0$&        &      &      &      &       &$P_0$ &        &      \\
\hline
\end{tabular}
\end{center}
\end{table}

The angular part of the Kahana-Ripka basis consists of the spherical 
harmonics, spin and isospin wave functions. 
Thus, introducing complex basis, one can rewrite the chiral fields 
in Eq.~(\ref{chiral_fields}) in the isospin space as  
\begin{eqnarray}
&& \bm{\tau} \cdot \hat{\bm{n}}_R=\tau_1 \hat{n}_1+\tau_2 \hat{n}_2+\tau_3 \hat{n}_3 \to
\tau_+ \hat{n}_-+\tau_- \hat{n}_++\tau_3 \hat{n}_3\,,\nonumber \\
&&\tau_\pm=\frac{1}{2}(\tau_1\pm i \tau_2)\,,  \nonumber \\
&&\hat{n}_+=\frac{2 R}{1+|R|^2}\,,~\hat{n}_-=\frac{2 R^*}{1+|R|^2}\,,~
\hat{n}_3=\frac{1-|R|^2}{1+|R|^2}\,.
\end{eqnarray}
$\hat{n}_{\pm}$ and $\hat{n}_3$ can be expanded by the spherical harmonics 
\begin{eqnarray}
&&\hat{n}(\theta,\phi)=\sum^{K_{max}}_{K=0}\sum^{K}_{M=-K}
a_{KM}Y_{KM}(\theta,\varphi)\,, \\
&&a_{KM}=\int \hat{n} (\theta,\varphi)Y^{*}(\theta,\varphi) \sin \theta d \theta d\phi\,.
\end{eqnarray}
Then we can perform the angular integration analytically 
with help of integral formula for the spherical harmonics given by 
\begin{eqnarray}
&&\int d\varphi \int \sin \theta d\theta Y_{l_1m_1}
(\theta,\varphi)Y_{l_2m_2}(\theta,\varphi)Y^{*}_{l_3m_3}(\theta,\varphi) \nonumber \\
&&\hspace{1cm}=\sqrt{\frac{(2l_1+1)(2l_2+1)}{4\pi (2l_3+1)}}C^{l_3 0}_{l_1 0 l_2 0}
C^{l_3 m_3}_{l_1 m_1 l_2 m_2} .
\end{eqnarray}

Since the chiral fields in Eq.~(\ref{chiral_fields}) is less symmetric than 
the $B=1$ hedgehog, the hamiltonian has no grand spin symmetry. 
As a result, the states with different grand spin couple strongly, and level 
splitting within the $K$ blocks occur. 
In Tables~\ref{tab:helement3}, \ref{tab:helement_nd3} are schematic pictures 
of the matrix elements ${\mathcal A}(K^{\prime}M^{\prime}, KM)$
and ${\mathcal B}(K^{\prime}M^{\prime}, KM)$ for $B=3$.
Although the size of the matrix becomes quite large, the functional space can be 
rearranged and reduced in size owing to the symmetry of the chiral 
fields. For $B=3$, the space is divided with four blocks for each parity. 

Numerically, we have to truncate the expansion of the size by cutoff 
$k_{\rm max}$ and  $K_{\rm max}$. In the tables, each element consists of a
 $4k_{\rm max}\times 4k_{\rm max}$ size of matrix. This block is spanned by 
$2\times \sum_{K=0}^{K_{\rm max}}(2K+1)/4$ for $B=3$ and 
hence the total matrix size is estimated as
\begin{eqnarray}
\sim \Bigl(K_{\rm max}(K_{\rm max}+2)/2\times 4k_{\rm max}\Bigr)^2.
\end{eqnarray}
The typical parameter values that we employ in our analysis 
are $k_{\rm max}=24$ and $K_{\rm max}=8$ for $B=3$, giving rather large matrix 
size $3800\times 3800$ approximately . 
In addition, our analysis is based on the self-consistent scheme and requires 
computational time much more. Obviously the situation gets worse for higher $B$. 
Therefore, in some cases such as $B=6, 17$, we oblige to cut
the size of the matrix, producing a few $\%$ uncertainty in results. 
In the next section we will 
discuss the convergence properties of our numerical 
computation in more detail. 

\begin{center}
\begin{figure}
\includegraphics[height=22cm, width=15cm]{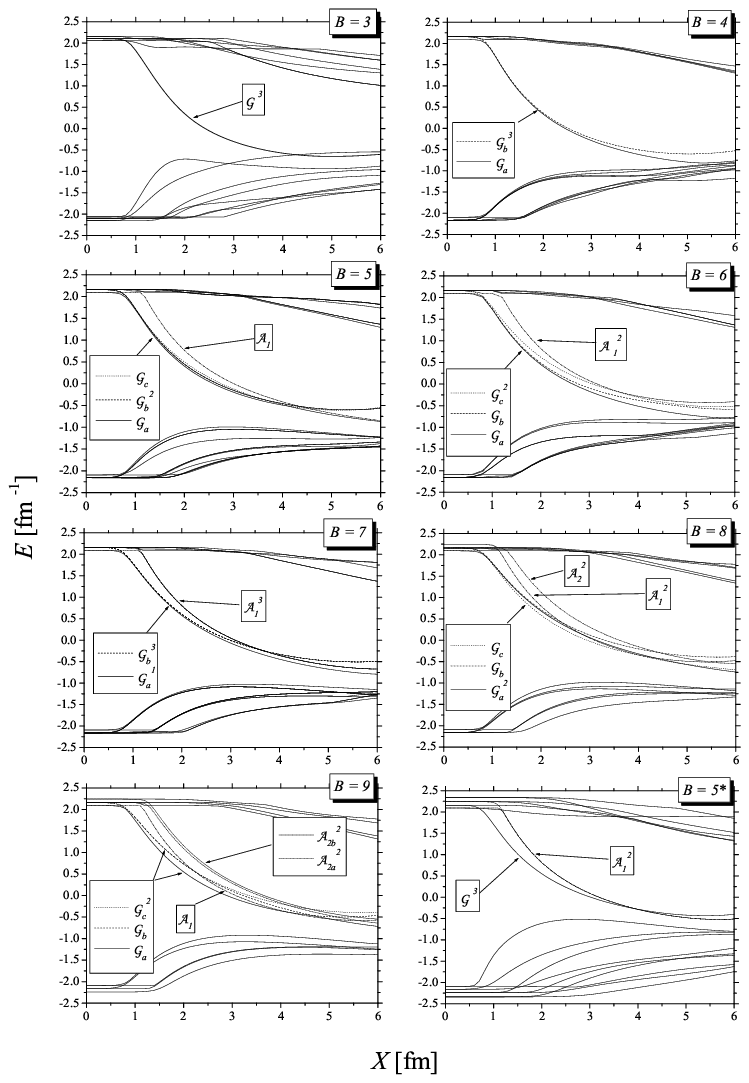}
\caption{\label{fig:elevels} Spectral flow of $B=3-9$ and $B=5$ excited state 
solutions with the occupation number.}
\end{figure}
\end{center}

\begin{figure}
\begin{center}
\includegraphics[height=8cm, width=10cm]{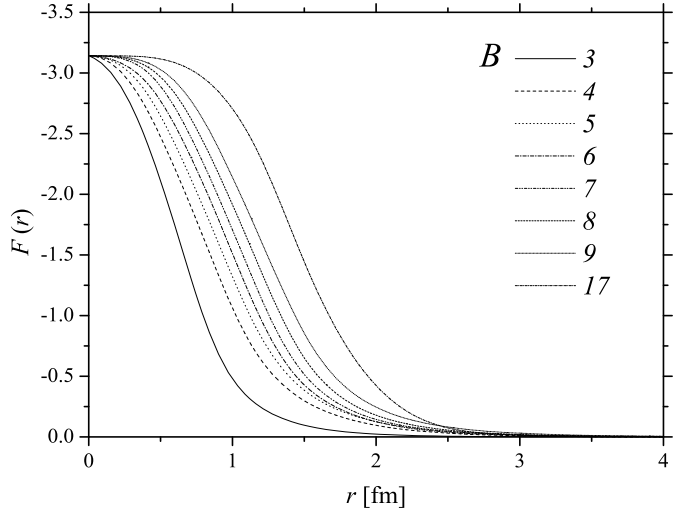}
\caption{\label{fig:profile} Self-consistent profile 
functions for $B=3-9,17$.}
\end{center}
\end{figure}




\begin{table}
\begin{center}
\caption{\label{tab:energy} Mass spectra for $B=1-9,17$ also for some excited states
$B=5^{*},9^{*}$(in MeV).
The data for $B=2$ are taken from Ref.~\cite{sawado00}.
The ratio of the mass $E_{\rm static}$ to $B\times E^{(B=1)}_{\rm static}$ are compared to 
that of the Skyrme model~\cite{manton98}.}
\begin{tabular}{lccccccccccccc} \hline
$B$ &\multicolumn{9}{c}{$E^{(i)}_{\rm val}$}&$E_{\rm field}$ 
&$E_{\rm static}$ &\multicolumn{2}{c}{$E_{\rm static}/B E^{(B=1)}_{\rm static}$}\\
    &\multicolumn{9}{c}{}  & && Ours & Skyrme\\ \hline
1    & 173 &     &     &     &     &&&  &  & 674  & 1192 & 1.00 & 1.00\\
2    & 173 & 173 &     &     &     &&&  &  & 1166 & 2204 & 0.92 & 0.95\\     
3    & 210 & 210 & 210 &     &     &&&  &  & 1633 & 3522 & 0.98 & 0.96\\ 
4    & 144 & 146 & 146 & 146 &     &&&  &  & 2628 & 4378 & 0.92 & 0.92\\
5    & 123 & 131 & 131 & 139 & 210 &&&  &  & 3265 & 5467 & 0.92 & 0.93\\ 
6    & 120 & 124 & 150 & 150 & 206 & 206 &&&     & 3740 & 6603 & 0.92 &0.92   \\
7    & 115 & 120 & 120 & 120 & 166 & 166 & 166 &&& 4554 & 7478 & 0.90 & 0.90 \\
8    & 97  & 97  & 115 & 120 & 139 & 139 & 203 & 203 && 5229 & 8565 & 0.90 & 0.91 \\
9    & 69  & 101 & 104 & 104 & 107 & 166 & 166 & 179 & 179 & 6046 & 9573 & 0.89& 0.90{\tiny 6} \\
17   & 83  & 95  & 95  & 95  & 153 & 156 & 157 & 173 & 175 &      &      &     &      \\
     & 177 & 178 & 179 & 192 & 194 & 194 & 196 & 196 &     & 10586& 18650& 0.93& 0.88\\
$5^{*}$& 157 & 157 & 157 & 232 & 232 &&&  &  & 2874 & 5680 & 0.95& 1.00\\ 
$9^{*}$& 99  & 105 & 105 & 121 & 142 & 142 & 210 & 210 & 210 & 5700 & 9742 & 0.91& 0.91 \\
\hline
\end{tabular}
\end{center}
\end{table}

\begin{figure}
\includegraphics[height=10cm, width=15cm]{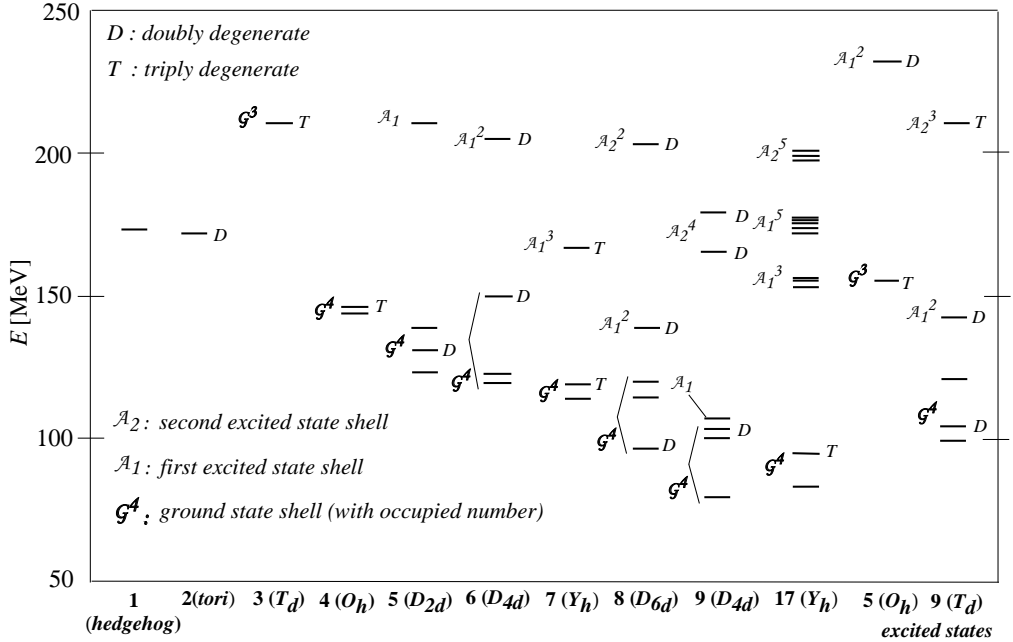}
\caption{\label{fig:spectrum} Valence quark spectra for $B=1-9,17$.}
\end{figure}

\begin{figure}
\includegraphics[height=22cm, width=15cm]{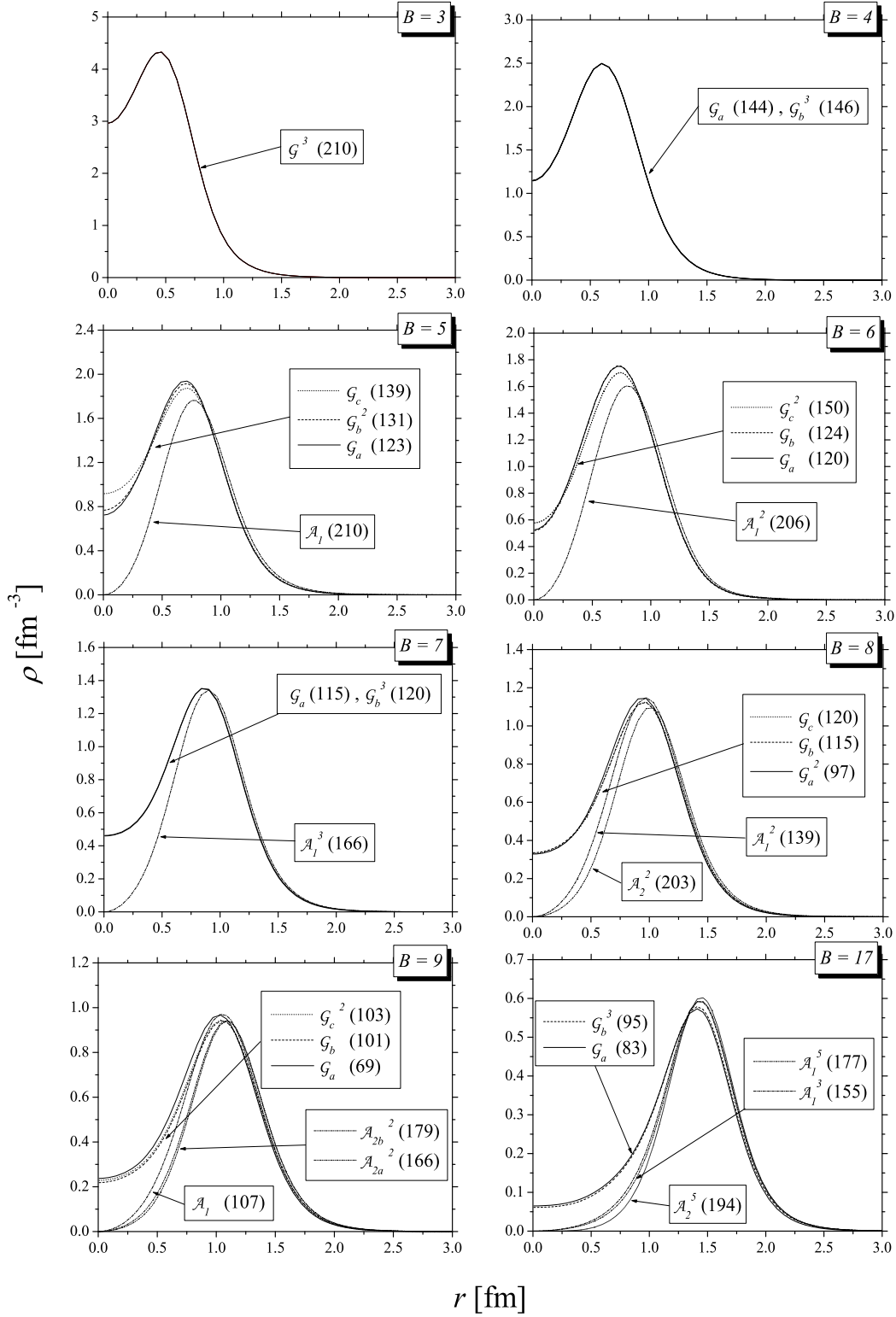}
\caption{\label{fig:bdrs} Angular averaged baryon number density 
of $i$th valence quarks $\rho^{(i)}(r)$ of $B=3-9,17$, 
with the occupation number and the eigenvalue (in MeV).}
\end{figure}

\begin{figure}
\includegraphics[height=6cm, width=7.5cm]{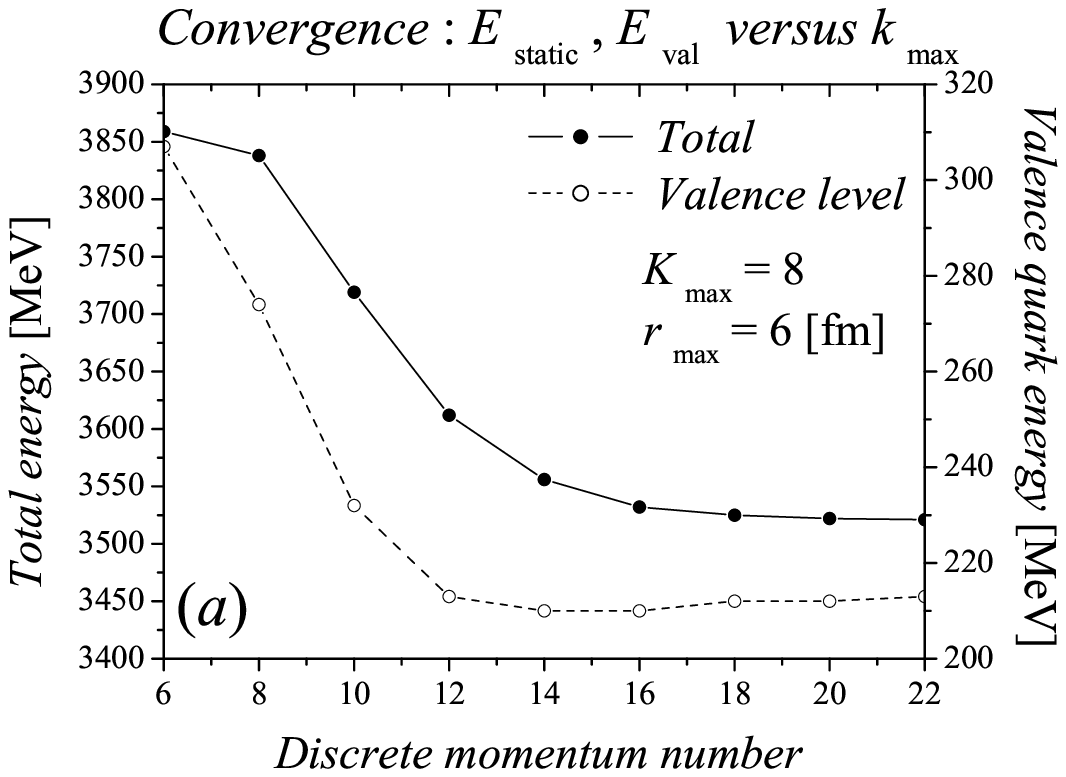}
\includegraphics[height=6cm, width=7.5cm]{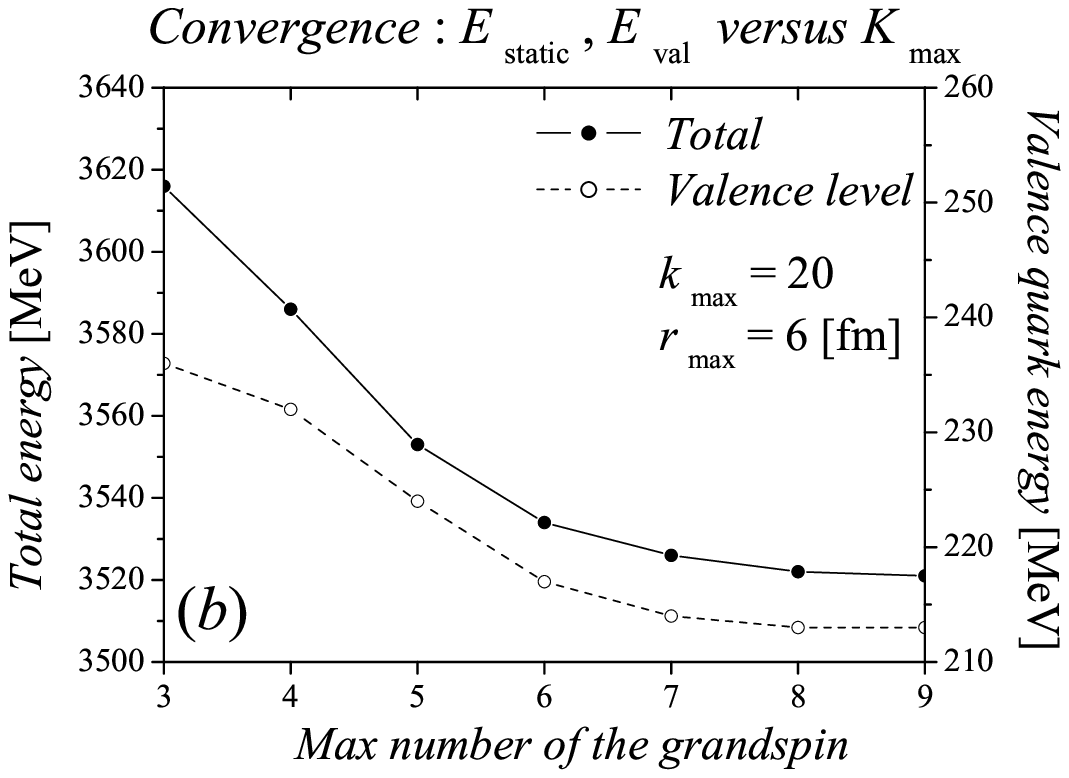}
\caption{\label{fig:convergence} The convergence properties of the 
solution with $B=3$. }
\end{figure}

\begin{figure}
\begin{center}
\includegraphics[height=6cm, width=7.5cm]{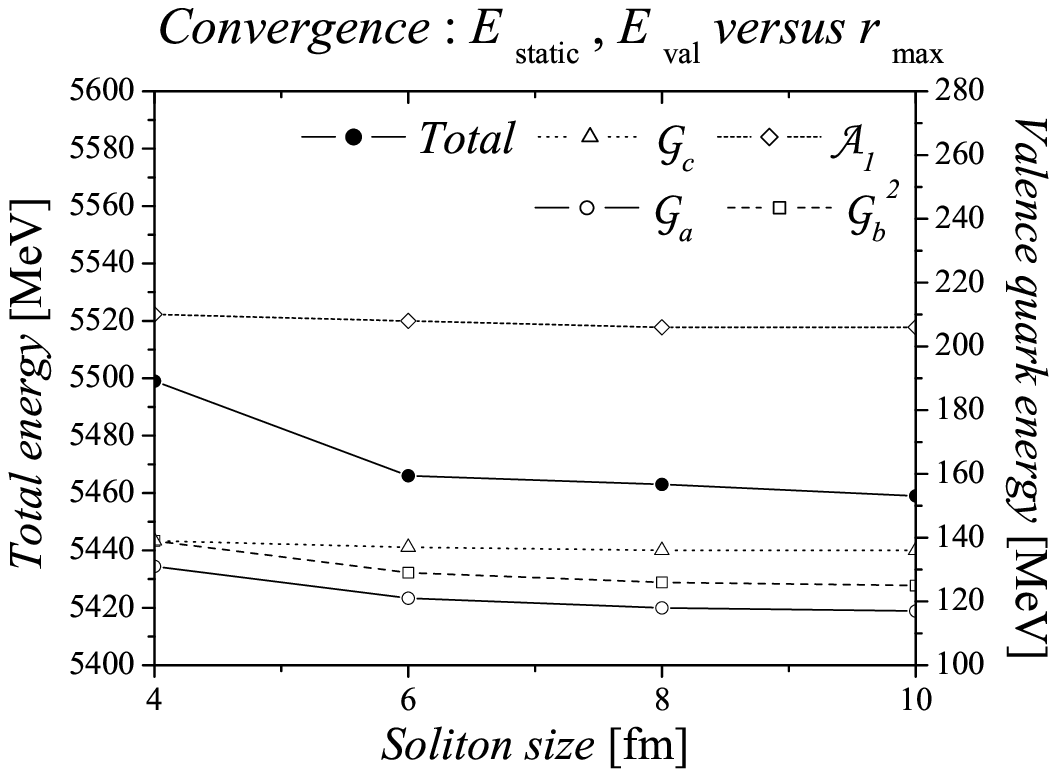}
\caption{\label{fig:convergence2} The convergence properties of the 
solution with $B=5$. $K_{\rm max}$ is fixed to be 8 and $m_{\rm max}$ 
is varied until convergence is attained.  }
\end{center}
\end{figure}

\section{\label{sec:level4}Results and discussions}
\subsection{Numerical results}
Let us first show the results of the spectral flow analysis.
For convenience we shall take 
\begin{eqnarray}
&&F(r)=-\pi+\pi r/X~~{\rm for}~~r< X \nonumber \\
&&\hspace{8.5mm}=0\hspace{2cm} {\rm otherwise}
\end{eqnarray} 
as a trial function for the profile function.   
In Fig.~\ref{fig:elevels} we show the spectral flows for $B=3-9$. 
As can be seen, the number of $B$ positive energy levels are 
diving into negative energy region and thus we obtain the 
baryon number $B$ soliton solutions.  
Putting the three quarks so as to be colour blind on each valence 
orbits as well as all on negative energy sea levels, we performed the 
self-consistent calculations. 
The profile functions for $B=3-9, 17$ are plotted in Fig.~\ref{fig:profile}. 
In Table \ref{tab:energy} are the results for the valence quark levels 
as well as the vacuum sea contributions. The valence quark spectra show 
various degenerate patterns depending on the background configuration. 
In Table \ref{tab:energy}, we also show the results of the ratio of 
the mass $E_{\rm static}$ to $B\times E^{B=1}_{\rm static}$, and 
comparison to that of the Skyrme model. The results are qualitatively in agreement.   

The valence quark spectra for various $B$ are 
shown in Fig.~\ref{fig:spectrum}. 
It is interesting that the results strongly suggest the existence 
of shell structure for the valence quarks. 
The spectra show (i) four fold degeneracy of the ground state 
labeled by ${\mathcal G}$ and various degenerate pattern for excited levels 
labeled by ${\mathcal A}_1, {\mathcal A}_2,\cdots$, 
(ii) a large energy gap between the ground state ${\mathcal G}$ and 
the first excited level ${\mathcal A}_1$. 
Small dispersions of the spectra are observed in the results. 
In some cases they are caused by the finite size effect of the basis 
(ex. $B=4$). Growing the size $r_{\rm max}$ and increasing the number of the basis, 
more accurate degeneracy will be attained.

All our solutions are local minima obtained under the rational 
map ansatz in Eq.~(\ref{chiral_fields}).  Of course there may exist other 
stable solutions with lower energy outside the ansatz.  We, however, suspect 
that the large degeneracy caused by the symmetry of the chiral fields would 
give a strong contribution to the minimization of the total energy. 

In Fig.~\ref{fig:bdrs} are the results of $\rho^{(i)}(r)$ for $B=3-9,17$.
The behaviour of the density near the origin confirms the existence of three shells 
(${\mathcal G},{\mathcal A}_1,{\mathcal A}_2$).
${\mathcal G}$ behaves like ``$S$-wave'' and others like ``$P$-,$D$-wave''
in a hydrogen-like atom.
However most of the densities are nearly on the same surface and very small 
(not zero) near the origin. 
The plateau in the density observed at the center of the nucleus~\cite{bohr} 
can not be attained in our solutions. Therefore one may need to employ the multi-shell 
ansatz~\cite{manton00} even in the case of light nuclei. 

Our numerical results depend on the basis parameters: 
radius $r_{\rm max}$, the number of discretized momenta $k_{\rm max}$, 
 and the maximum value of the grandspin $K_{\rm max}$. 
In Fig.~\ref{fig:convergence} we demonstrate the accuracy of our
numerical computation for $B=3$ with $r_{\rm max}=6$ [fm].  The convergence of the 
solution and the valence energy $E_{\rm static}$, $E_{\rm val}$ are shown  
as a function of $(a)$ $k_{\rm max}$ with fixed $K_{\rm max}$ and 
$(b)$ $K_{\rm max}$ with fixed $k_{\rm max}$. This confirms that 
for $K_{\rm max}=8$ and $ k_{\rm max}=20$ the solution with $B=3$ 
is sufficiently converging.

As stated above, dispersions appear in the spectra, and in  
some cases it will be eliminated by taking a larger size of 
the spherical box and increasing the number of basis.  
However, for most of the solutions obtained here, these dispersions  
can not be ascribed to numerical errors. 
Let us examine the relation of the energy and the soliton 
size $r_{\rm max}$ for $B=5$ (see Fig.~\ref{fig:convergence2}). 
For each value of $r_{\rm max}$, we employ sufficiently large 
number of the basis to attain convergence. As can be seen, the dispersion 
does not disappear even for larger value of $r_{\rm max}$.    
We therefore conclude that these dispersions are not due to numerical errors nor
uncertainty but are the inherent feature of the solutions.  

\subsection{Symmetry and the degeneracy of the quarks}
The bunch of valence spectra due to the potential with discrete symmetries 
has been observed in the study of heavier nuclear systems. In Ref.~\cite{dudek}, 
the valence spectra are highly degenerate because the deformation 
of the spherically symmetric shell produces large shell gaps.  
Thus the nuclei can be considered to be more stable than the spherical one. 
As discussed in Ref.~\cite{manton98}, the group theory should predict 
the level structure of pion fluctuations. However, our problem is 
more complicated due to the presence of quarks. 
Before discussing it in detail, let us show how the shell deformation 
is related to the degeneracy of the spectrum. 

In general, if an eigenequation given by 
\begin{eqnarray}
H\psi_\mu=E_\mu\psi_\mu
\end{eqnarray}
is invariant under a symmetric operation $\hat{R}$ $\in$ $g$, 
the equation transforms as
\begin{eqnarray}
\hat{R}H\psi_\mu=H(\hat{R}\psi_\mu)=E_\mu\psi_\mu\,.
\end{eqnarray}
Therefore the states \{$\psi_\mu, \hat{R}\psi_\mu$\} are degenerate in energy with $E_\mu$. 
The set of $d_\mu$ eigenfunctions \{$\psi_i^{(\mu)}$\}$(i=1,\dots,d_\mu)$ belonging to a given 
eigenvalue $E_\mu$ will provide the 
basis for an irreducible representation of the group $g$ of the hamiltonian
\cite{hamermesh}:
\begin{eqnarray}
\hat{R}\psi_j^{(\mu)}=\sum_i\psi_i^{(\mu)} D^{(\mu)}_{ij}(\hat{R})\,.
\end{eqnarray}

The operator $\hat{R}$ are constructed as follows.
If chiral fields have some particular point group symmetry i.e., 
$U(x')=\hat{G}(\hat{a})U(x)\hat{G}(\hat{a})^\dagger~(\hat{G}(\hat{a})\in SU(2)_I$, 
and $\hat{a}$ denotes the matrix of the following spatial rotation $a^\nu_\mu$),
the Dirac equation is invariant under the Lorentz transformation 
\begin{eqnarray}
(x')^\nu=a^\nu_\mu x^\mu~~
{\rm or}~~x'=\hat{a} x 
\end{eqnarray}
with
\begin{eqnarray}
x'=
\left(\begin{array}{c} 
	~t \\
	~\bm{x}' 
	\end{array}
	\right)\,,
~~\hat{a}=
\left(\begin{array}{cc} 
	~1&~0~ \\
	~0&~\hat{\bm{a}} 
	\end{array}
	\right)  \,,
~~x=
\left(\begin{array}{c} 
	~t \\
	~\bm{x}
	\end{array}
	\right)\,,
\end{eqnarray}
accompanying a corresponding iso-rotation
\begin{eqnarray}
(i\gamma^\mu \partial_\mu-MU^{\gamma_5}(x))\psi(x)=0 
\Rightarrow (i\gamma^\nu \partial'_\nu-MU^{\gamma_5}(x'))\psi'(x')=0\,,
\end{eqnarray}
with 
\begin{eqnarray}
\psi'(x')=(\hat{S}(\hat{a})\times \hat{G}(\hat{a}))\psi(x)\,,
\end{eqnarray}
where $\hat{S}(\hat{a})$ is a $4\times 4$ matrix which is a function of the 
parameters of the Lorentz transformation $\hat{a}$, satisfying 
$a^{\nu}_{\mu}\gamma^\mu=\hat{S}^{-1}\gamma^\nu \hat{S}$.
The operator $\hat{R}$ corresponding to this rotation is thus defined by
\begin{eqnarray}
\psi'(x)\equiv_{def} \hat{R}\psi(x)=\hat{S}(\hat{a})\times \hat{G}(\hat{a})\psi(\hat{a}^{-1}x)\,.
\end{eqnarray}
One can easily check that $R$ commutes with the 
hamiltonian in Eq. (\ref{hamiltonian}).
Constructing $\hat{R}$ for each symmetry of the hamiltonian, one should be able to 
deduce the degeneracy structure of the spectra occurring in the valence level.  
The construction of $\hat{R}$ in the case of $B=3$ is shown in the appendix B, which 
confirms that the valence level has triply degenerate spectra. 

Our numerical results indicate that the winding number strongly couple 
the elements with different $K$ and hence correlated valence spectra occur 
(see Table \ref{tab:helement3}-\ref{tab:helement_nd3}). 
As can be seen from the operator $K$ of the $B=2$, the degeneracy of the valence spectra 
are determined by the shape deformation (symmetry) as well as the winding number  
of the chiral fields~\cite{sawado98}. The four-fold degeneracy of the lowest states 
may be ascribed to the chiral symmetry $SU(2)_L\times SU(2)_R$ of the hamiltonian. 
The degenerate structure for $B\ge 3$ will be well understood if symmetric operators 
of the hamiltonian which consist of the angular momentum, spin, isospin and winding 
number, are explicitly constructed. 

\section{\label{sec:level5}Summary}
In this paper we investigated the multi-soliton solutions in the chiral quark 
soliton model, using the rational map ansatz as a background chiral fields for quarks. 
The chiral fields with multi-winding number have particular discrete symmetries 
and it was shown that the baryon densities inherit the same discrete symmetries 
as the chiral fields.  For the quark levels we observed 
various degenerate bound spectra depending on the background of chiral field configurations. 
Evaluating the radial component of the baryon density, shell-like structure of the 
valence quark spectra was also observed. 
The group theory should predict these level structures resulting from the symmetry
of the background potential. In fact the degeneracy of the valence spectra 
are determined by the winding number of the chiral fields as well as the shape 
deformation (symmetry) of solitons. The four-fold degeneracy of the lowest states may be 
ascribed to the chiral symmetry $SU(2)_L\times SU(2)_R$ of the hamiltonian. 
To get better understanding of the relation between the quark level structure and 
the winding number or the shape deformation, further analysis will be worth to be done 
in future.  

\ack
We are grateful to N.S.Manton for encouraging us to work on this subject 
and many useful comments. We also thank S.Krusch for his careful reading
and comments about our manuscript. We are indebted to M.Kawabata and K.Saito 
for their help of numerical computations. 

\appendix
\section{\label{sec:levela}The rational maps}
In this appendix, we present the explicit forms of the rational maps which 
we used in our analysis as a background chiral fields 
(from Refs. \cite{manton98,battye01}). 
\begin{eqnarray}
&&R_3=\frac{\sqrt{3}iz^{2}-1}{z(z^{2}-\sqrt{3}i)} \,,\nonumber\\
&&R_4=\frac{z^4+2\sqrt{3}iz^2+1}{z^4-2\sqrt{3}iz^2+1} \,,\nonumber \\
&&R_5=\frac{z(z^4+3.94z^2+3.07)}{3.07z^4-3.94z^2+1} \,, \nonumber \\
&&R_6=\frac{z^4+0.16 i}{z^2(0.16z^4 i+1)} \,,\nonumber \\
&&R_7=\frac{7/\sqrt{5}z^6-7z^4-7/\sqrt{5}z^2-1}{z(z^6+7/\sqrt{5}z^4+7z^2-7/\sqrt{5})} \,,\nonumber \\
&&R_8=\frac{z^6-0.14}{z^2(0.14z^6+1)}\,, \nonumber \\
&&R_9=\frac{z(-3.38-11.19iz^4+z^8)}{1-11.19iz^4-3.38z^8}\nonumber \\
&&R_5^{*}=\frac{z(z^4-5)}{-5z^4+1} \,,\nonumber \\
&&R_9^{*}=\frac{5i\sqrt{3}z^6-9z^4+3i\sqrt{3}z^2+1-1.98z^2(z^6-i\sqrt{3}z^4-z^2+i\sqrt{3})}
{z^3(-z^6-3i\sqrt{3}z^4+9z^2-5i\sqrt{3}-1.98z(-i\sqrt{3}z^6+z^4+i\sqrt{3}z^2-1))}\,, \nonumber \\
&&R_{17}=\frac{17z^{15}-187z^{10}+119z^5-1}{z^2(z^{15}+119z^{10}+187z^5+17)}\,.
\end{eqnarray}

\section{\label{sec:levelb}The Lorentz transformation with the chiral fields}
In this appendix, we briefly show the evaluation of the rotation operator $\hat{R}$ for $B=3$ tetrahedron
and also present the results of the transformation law for the numerical basis~(\ref{kahana_ripka}). 
The $B=3$ tetrahedral soliton is characterized by two symmetry operations~\cite{manton98}:
$Z_2\times Z_2$ and $T_d$. The Lorentz transformation operators $\hat{S}$ and the operators for 
the chiral fields $\{g,h\}$ corresponding to the symmetry operations $(x')^\nu=a^\nu_\mu x^\mu$ are given by
\begin{eqnarray}
&&Z_2\times Z_2: \nonumber \\
&&\hspace{1cm}(a_g)^\nu_\mu=
\left(\begin{array}{cccc} 
	~1&~0&~0&~0 \\
	~0&-1&~0&~0 \\
	~0&~0&~1&~0 \\
	~0&~0&~0&-1 \\
	\end{array}
	\right)~~\Rightarrow~~
\hat{S}_g=i\gamma^0\gamma^5\gamma^2\,,  \nonumber \\
&&\hspace{1cm}U(\hat{a}_g x)=\hat{g}U(x)\hat{g}^{\dagger}\,,~~\hat{g}=\exp[-i\frac{\pi}{2}\tau_2]\,.~~\nonumber \\
&&T_d: \nonumber \\
&&\hspace{1cm}(a_h)^\nu_\mu=
\left(\begin{array}{cccc} 
	~1&~0&~0&~0 \\
	~0&~0&~1&~0 \\
	~0&~0&~0&~1 \\
	~0&~1&~0&~0 \\
	\end{array}
	\right)  
\Rightarrow
\hat{S}_h=\exp[i\frac{\pi}{3}\frac{1}{\sqrt{3}}(\sigma_{23}+\sigma_{31}+\sigma_{12})]\,,
\nonumber \\
&&\hspace{1cm}U(\hat{a}_hx)=\hat{h}U(x)\hat{h}^{\dagger}\,,~~
\hat{h}=\exp[-i\frac{\pi}{3}\frac{1}{\sqrt{3}}(\tau_1+\tau_2+\tau_3)]\,.
\nonumber
\end{eqnarray}
The $\hat{R}$ is defined by the direct product of these rotation operators toghther
with the inverse spatial rotation for the spinor, that is 
\begin{eqnarray}
&&\hat{R}_g\psi(x)\equiv (\hat{S}_g\times \hat{g})\psi(\hat{a}_g^{-1}x)\,, \\
&&\hat{R}_h\psi(x)\equiv (\hat{S}_h\times \hat{h})\psi(\hat{a}_h^{-1}x)\,.
\end{eqnarray} 
We apply these operators to the Kahana-Ripka basis $\phi\equiv\{u,v\}$ and finally 
obtain the following transformation law:
\begin{eqnarray}
&&\hat{R}_g\phi_{KM}=(-1)^{K-M}\phi_{K-M}\,. \\
&&\hat{R}_h\phi_{00}=\phi_{00}\,. \\
&&\hat{R}_g\left(
\begin{array}{c}
\phi_a \\\phi_b \\\phi_c
\end{array}
\right)=
\left(
\begin{array}{ccc}
~0&-1&~0 \\
~0&~0&-1 \\
~1&~0&~0  \\
\end{array}
\right)
\left(
\begin{array}{c}
\phi_a \\\phi_b \\\phi_c
\end{array}
\right)\,, \\
&&\phi_a\equiv\frac{1}{\sqrt{2}}(\phi_{11}+\phi_{1-1})\,,~~
\phi_b\equiv i\phi_{10}\,,~~
\phi_c\equiv \frac{i}{\sqrt{2}}(\phi_{11}-\phi_{1-1})\,.
\nonumber \\
&&\hat{R}_h\left(
\begin{array}{c}
\phi_\xi \\\phi_\eta \\\phi_\zeta \\\phi_u \\\phi_v
\end{array}
\right)=
\left(
\begin{array}{ccc|cc}
~0&-1&~0~& 0& 0\\
~0&~0&~1~& 0& 0\\
-1&~0&~0~& 0& 0\\ \hline
 0& 0& 0&-\frac{1}{2}&~\frac{\sqrt{3}}{2}  \\
 0& 0& 0&-\frac{\sqrt{3}}{2} & -\frac{1}{2}\\
\end{array}
\right)
\left(
\begin{array}{c}
\phi_\xi \\\phi_\eta \\\phi_\zeta \\\phi_u \\\phi_v
\end{array}
\right)\,. \\
&&\phi_\xi=\frac{i}{\sqrt{2}}(\phi_{21}+\phi_{2-1})\,,~~
\phi_\eta=\frac{1}{\sqrt{2}}(\phi_{21}-\phi_{2-1})\,, \nonumber \\
&&\phi_\zeta=\frac{i}{\sqrt{2}}(\phi_{22}-\phi_{2-2})\,,~~
\phi_u=\phi_{20}\,,~~
\phi_v=\frac{1}{\sqrt{2}}(\phi_{22}+\phi_{2-2})\,. \nonumber 
\end{eqnarray}
Thus we confirm that the $B=3$ tetrahedron exhibits triply degenerate spectra.


\begin{thebibliography}{99}
\bibitem{diakonov88}
D. I. Diakonov, V. Yu. Petrov, and P. V. Pobylitsa, 
Nucl. Phys. {\bf B306,} 809 (1988).
\bibitem{reinhardt88}
H. Reinhardt and R. W\"unsch , Phys. Lett. {\bf B215,} 577 (1988).
\bibitem{meissner89}
Th. Meissner, F. Gr\"ummer, and K. Goeke, Phys. Lett. 
{\bf B227,} 296 (1989).
\bibitem{report} For detailed reviews of the model see: \\
R.\ Alkofer, H.\ Reinhardt and H.\ Weigel, Phys.\ Rept.\ {\bf 265}, 139 (1996);\\
 Chr.\ V.\ Christov, A.\ Blotz, H.-C.Kim, P.\ Pobylitsa, T.\ Watabe, Th.\ Meissner, 
E.\ Ruiz Arriola, K.\ Goeke, Prog.\ Part.\ Nucl.\ Phys.\ {\bf 37}, 91 (1996).
\bibitem{wakamatsu91}
M. Wakamatsu and H. Yoshiki, Nucl. Phys. {\bf A524}, 561 (1991).
\bibitem{sawado98}
N. Sawado and S. Oryu, Phys. Rev. {\bf C58}, R3046 (1998).
\bibitem{sawado00}
N. Sawado, Phys. Rev. {\bf C61}, 65206 (2000).
\bibitem{sawado02}
N. Sawado, Phys. Lett. {\bf B524}, 289 (2002).
\bibitem{braaten90}
E. Braaten, S. Townsend and L. Carson, 
Phys. Lett. {\bf B235}, 147 (1990).
\bibitem{sutcliffe97}
R. A. Battye and P. M. Sutcliffe, 
Phys. Rev. Lett. {\bf 79}, 363 (1997).
\bibitem{manton98}
C. J. Houghton, N. S. Manton and P. M. Sutcliffe, 
Nucl. Phys. {\bf B510}, 507 (1998).
\bibitem{sawado02t}
N. Sawado and N. Shiiki,
Phys. Rev. {\bf D66}, 011501 (2002).
\bibitem{dhar85}
A. Dhar, R. Shankar, S. R. Wadia, 
Phys. Rev. {\bf D31}, 3256 (1985).
\bibitem{ebert86}
D. Ebert, H. Reinhardt, 
Nucl.Phys. {\bf B271}, 188 (1986).
\bibitem{aitchison85}
I. Aitchison, C. Fraser, E. Tudor and J. Zuk,  
Phys. Lett. {\bf B165}, 162 (1985).
\bibitem{kahana84}
S. Kahana, G. Ripka, and V. Soni, Nucl. Phys. {\bf A415,} 
351 (1984); S. Kahana and G. Ripka, \textit{ibid}, {\bf A429,} 462 (1984).
\bibitem{balachandran98}
A. P. Balachandran and S. Vaidya,
Int. J. Mod. Phys. {\bf A14}, 445 (1999),
{\it hep-th/9803125}.
\bibitem{schwinger51}
J. Schwinger, Phys. Rev. {\bf 82}, 664 (1951).
\bibitem{bransden}
B. H. Bransden and C. J. Joachain, {\it Introduction to Quantum 
Mechanics} (Longman Scientific Technical, 1989)
\bibitem{bohr}
A. Bohr and B. Mottelson, {\it Nuclear structure, Vol.I}
(World Scientific Publishing Co. Pte. Ltd, Singapore, 1998).
\bibitem{manton00}
N. S. Manton and B. M. A. G. Piette, 
Prog. Math. {\bf 201}, 469 (2001);{\it hep-th/0008110}.
\bibitem{dudek}
J. Dudek, A. Go\'zd\'z, N. Schunck, and M. Mi\'skiewicz, 
Phys. Rev. Lett. {\bf 88}, 252502 (2002).
\bibitem{hamermesh}
M. Hamermesh, {\it Group theory and its application to physical
problems} (Dover Publications, Inc., New York, 1989).
\bibitem{battye01}
R. A. Battye and P. M. Sutcliffe, 
Rev. Math. Phys. {\bf 14}, 29, (2002).
\end{thebibliography}
\end{document}